\documentclass[11pt]{extarticle}

\usepackage{amsfonts}
\usepackage{graphicx} 
\usepackage{color}
\usepackage{subfigure}
\usepackage{amsmath} 
\usepackage{amssymb} 
\usepackage{mathrsfs}
\usepackage{alltt}
\usepackage{cite}
\usepackage{multicol}
\usepackage{multirow}
\usepackage{threeparttable}
\usepackage{float} 

\usepackage{geometry} 
\def\bm{\mathbf}
\def\bm{\boldsymbol}
\newtheorem{Theo}{Theorem}[section]
\newcommand{\bo}{\begin{Theo}}
\newcommand{\eo}{\end{Theo}}
\newtheorem{Lem}{Lemma}
\newcommand{\bl}{\begin{Lem}}
\newcommand{\el}{\end{Lem}}

\title{Functional linear regression for partially observed functional data}

\author{Yafei Wang$^{a,b}$, Tingyu Lai$^{a}$, Bei Jiang$^{b}$, Linglong Kong$^{b}$\footnote{Corresponding author: lkong@ualberta.ca}, Zhongzhan Zhang$^{a}$\footnote{Corresponding author: zzhang@bjut.edu.cn}\\
$^a$Beijing University of Technology, Beijing, 100124, China,\\
$^b$University of Alberta, Edmonton, T6G 2G1, Canada.
}
\date{}

\begin{document}

\maketitle

\begin{abstract}
In the functional linear regression model, many methods have been proposed and studied to estimate the slope function while the functional predictor was observed in the entire domain. However, works on functional linear regression models with partially observed trajectories have received less attention. In this paper, to fill the literature gap we consider the scenario where individual functional predictor may be observed only on part of the domain. Depending on whether measurement error is presented in functional predictors, two methods are developed, one is based on linear functionals of the observed part of the trajectory and the other one uses conditional principal component scores. We establish the asymptotic properties of the two proposed methods. Finite sample simulations are conducted to verify their performance. Diffusion tensor imaging (DTI) data from Alzheimer's Disease Neuroimaging Initiative (ADNI) study is analyzed.

\end{abstract}

{\bf Keywords:} {Functional linear model; Partially observed functional data; Principal components; Measurement error; ADNI.}

\section{Introduction}
\label{}
With the advance in technology, it is increasingly common to encounter data that are functions or curves in nature (see \cite{ramsay2005functional}). Functional linear regression models provide a framework for modeling the dynamic relationship between response and functional predictors, which was first introduced by \cite{ramsay1991some}. One of the primary goals for functional linear model (FLM) is to get an estimator of functional coefficient. And many procedures have been proposed to approximate functional coefficient, for example, functional principal component analysis (FPCA) based approaches (\cite{cardot1999functional}, \cite{hall2007methodology}, \cite{yao2005bfunctional}), spline-based approaches (\cite{crambes2009smoothing}, \cite{marx1999generalized}), wavelet-based approaches (\cite{zhao2012wavelet}, \cite{wang2019wavelet}), and others. 
We refer to \cite{morris2015functional} and \cite{reiss2017methods} for more informative and extensive reviews on such functional linear models.

Among the different based methods in functional data analysis, FPCA based approaches for capturing the information of covariates are popular (\cite{hall2006properties}, \cite{che2017trajectory}). In the setting where trajectories are observed on dense and regular grid on the entire domain, the existing works can be found in \cite{besse1986principal}, \cite{rice1991estimating}, \cite{cardot1999functional}, \cite{shin2009partial}, \cite{horvath2012inference}, to name a few. \cite{yao2005afunctional} emphasizes the case where the functional predictors are observed with irregularly sparse measurements which is often referred to as sparse functional data, and proposes a nonparametric method to perform FPCA.  For general review on FPCA, see \cite{shang2014survey}. In this paper, we prefer to use FPCA method to get an estimator of the functional coefficient.

Sparse functional data addresses the case where each trajectory is observed at a small number of points that are distributed randomly on the domain which is different from the partially observed functional data (or incomplete or fragmentary functional data) which was first introduced in \cite{liebl2013modeling}. Partially observed functional data addresses each trajectory is observed at points that cover a subset of the domain in such a way that trajectories can be reasonably treated as fragments of curves (\cite{delaigle2016approximating}) that has great implication in applications, such as in biomedicine, economics (see \cite{kraus2015components}, \cite{kneip2020optimal}). Considering the partially observed functional data can be treated as missing data for functional curves over the domain, two missing mechanisms are introduced in the existing works: one is missing completely at random (MCAR), that is, the missing data mechanism is independent from other stochastic components (\cite{delaigle2016approximating}, \cite{goldberg2014predicting}); the other one is the missing mechanism in which depends on systematic strategies, such as missing parts of the trajectories only occur at the upper interval of the domain (see \cite{liebl2019csda}). In the setting of MCAR, \cite{delaigle2016approximating}, \cite{goldberg2014predicting} and \cite{kraus2015components} address the problem for recovering the missing parts of trajectories. \cite{kraus2015components} and \cite{kneip2020optimal} model the functional principal component (FPC) scores of an incomplete trajectory. In the scenario where missing data mechanism depends on systematic strategies, \cite{liebl2019csda} establishes estimators for the mean and the covariance function of the incomplete functional data via the fundamental theorem of calculus. To the best of our knowledge, no work exists focusing on estimating functional coefficient of FLM with partially observed trajectories.

In this paper, we address the problem of getting an estimator of functional coefficient for the case of partially observed functional data without and with measurement error. In the scenario that trajectories observed without measurement error, instead of deleting the incomplete trajectories, we get estimators of FPC scores for each incomplete trajectory by modeling it as linear functionals of the observed parts of that trajectory. In the setting where trajectories observed with measurement error, we use local linear smoother methods to estimate mean and covariance function of the functional predictor, followed by getting FPC scores via conditional expectation.

The contributions of this paper are as follows. First, we extend FLM approach to partially observed functional data without measurement error, which leads to an improved estimator for functional coefficient comparing with the one obtained through deleting the incomplete trajectories for given dataset. Second, we develop an estimate method for functional coefficient in FLM for incomplete trajectories with measurement error. We illustrate its usefulness by comparing with another two methods: one is based on integration method to get the FPC scores of the functional predictor instead of using conditional expectations; the other estimator is obtained by ignoring the measurement error of the trajectories in the dataset. Third, in both scenarios, we obtain the rate of convergence for the proposed estimators. Overall, the methodological and numerical developments in this paper can provide a practically useful way in analyzing FLM with partially observed functional data.

The rest of this paper is organized as follows. In Section 2, we introduce functional linear models. In Section 3.1, we develop an estimator for functional coefficient with incomplete trajectories observed without measurement error, and establish theoretical properties for the proposed estimator. An estimator and theoretical properties in the scenario that incomplete trajectories observed with measurement error is introduced in Section 3.2.  Section 4 illustrates the finite sample performance of our proposed estimators through simulation studies, followed by a real data analysis in Section 5. Discussion is presented in Section 6. Proofs of theorems are given in the Appendix.

\section{Functional Linear Model}

Consider a functional linear model, in which the scalar response $Y_i$ is linearly related to the functional covariate $X_i$,
\begin{equation*}
Y_i= \alpha +  \int_\mathcal{T}\gamma(t)X_i(t)dt+\epsilon_i, \tag{1} \label{mainmodel}
\end{equation*}
where $\alpha$ is the intercept,  $\{X_i(t): t\in\mathcal{T},i=1,\ldots,n\}$ are the functional predictors, sampled from the stochastic process
$\{X(t):t\in\mathcal{T}\}$ with mean function $\mu$, domain $\mathcal{T}$ is bounded and closed,
$\gamma$ is the slope function to be estimated, $\epsilon_i$ are random errors
satisfying
$\text E[\epsilon_i]=0$, $\text E [\epsilon_i^2]=\sigma^2<\infty$. We can easily get an estimator of intercept once we get an estimator of $\gamma$. So we focus on estimating $\gamma$ in the following (\cite{hall2007methodology}).
Let $\langle\cdot, \cdot\rangle$, $||\cdot||$ be the inner product and norm on $L^2(\mathcal{T})$, the set of all square integrable functions on $\mathcal{T}$, with $\langle f, g \rangle = \int_{\mathcal{T}}f(t)g(t)\text{dt}$, $\|f\| = \langle f,f\rangle^{1/2}$ for any $f,g\in L^2(\mathcal{T})$. 

We first recall the method FPCA in estimating the slope function for model \eqref{mainmodel} with the functional predictor $X_i$ observed on the entire domain $\mathcal{T}$. For the stochastic process $X\in L^2(\mathcal{T})$, denote its mean function as $\mu$: $\mu = \text{E}(X)$, and its covariance function as $c_X(s, t)$: $c_X(s, t) = \text{cov}(X(s), X(t))$. Assume  $c_X$ is continuous on $\mathcal{T}\times\mathcal{T}$. The expression $c_X(s,t)=\sum_{j=1}^\infty \lambda_j\phi_j(s)\phi_j(t)$ exists by the Mercer Lemma (\cite{riesz1955b}), where $\lambda_1>\lambda_2>\cdots>0$; $\phi_1,\phi_2,\cdots$ are the eigenvalue sequence and the continuous orthonormal eigenfunction sequence of the linear operator $C_X$: $(C_X\phi)(\cdot) = \int_{\mathcal{T}} c_X(\cdot,t)\phi(t)\text{dt}$, $\phi\in L^2(\mathcal{T})$, with the kernel $c_X$. On the other hand, by the Karhunen-Lo\`{e}ve (K-L) expansion, one has
$
X_i(t)=\sum_{j=1}^\infty U_{ij}\phi_j(t),
$
where the random variables $U_{ij}=\langle X_i - \mu,\phi_j\rangle$ are uncorrelated with $\text E [U_{ij}]=0,\,\text E [U_{ij}^2]=\lambda_j$,
and
$
\gamma(t)=\sum_{j=1}^\infty\gamma_j\phi_j(t)
$
with $\gamma_j=\langle \gamma,\phi_j\rangle$. 

The full model \eqref{mainmodel} is then equivalent to
$
Y_i - \text{E}Y_i= \sum_{j=1}^\infty \gamma_j U_{ij}+\epsilon_i
$
based on K-L expansion, which can be approximated by $\sum_{j=1}^m\gamma_j {U}_{ij}+\epsilon_i$ by using the first $m$ terms. To simplify notations, we assume that $\{Y_i,i=1,\cdots,n\}$ are centered.  
Let $\bm Y=(Y_1,\cdots,Y_n)^T,\, {\bm \gamma}=(\gamma_1,\cdots,\gamma_m)^T$, $\hat{\mu}$ be an estimator of $\mu$, $\{\hat{\lambda}_j\}$ and $\{\hat{\phi}_j\}$ be estimators of the sequence $\{{\lambda}_j\}$ and $\{{\phi}_j\}$ with $\hat{\lambda}_1>\hat{\lambda}_2>\cdots>0$.
The least square estimator $\hat{\bm{{\gamma}}}$ is then given as
\begin{align*}
\hat{\bm{{\gamma}}}=({\hat{\bm U}_m}^T\hat{\bm U}_m)^{-1}\hat{\bm U}_m \bm Y,\tag{2}\label{gammasep}
\end{align*}
provided that $({\hat{\bm U}_m}^T\hat {\bm U}_m)^{-1}$ exists with $\hat{U}_{ij}=\langle X_i - \hat{\mu},\hat{\phi}_j\rangle$, $\hat{\bm U}_m=(\hat{U}_{ij})_{\substack{i=1,\cdots,n;\\ j=1,\cdots,m}}$. Moreover, for the estimator $\hat{\gamma}_j, j=1,\cdots,m$, it has the equivalent form as
\begin{equation*}
\hat{\gamma}_j = \hat{\lambda}^{-1}_{j}\left\langle n^{-1}\sum_{i=1}^{n} (Y_i-\bar{Y}_0)(X_i-\hat{\mu}), \hat\phi_{j}\right\rangle. 
\end{equation*}
Consequently, an estimator of $\gamma$ is given by
\begin{equation*}
\hat \gamma(t)=\sum_{j=1}^m \hat\gamma_j\hat\phi_{j}(t). \tag{3} \label{gammahat}
\end{equation*}

The number $m$ of included eigenfunctions is chosen by fraction of variance explained criterion in practice (\cite{james2000principal}): $m=\text{min}\{k:\sum_{l=1}^{k}\hat{\lambda}_l/\sum_{l=1}^{n}\hat{\lambda}_i\geq R \}$, with a given threshold $R$. For the asymptotic analysis, we assume $m$ depends on sample size $n$ such that $m \rightarrow\infty$ as $n\rightarrow\infty$.

\section{Estimation Methods}\label{method}
The above analysis is based on the assumption the functional predictor is observed on the entire domain. We now consider the scenario that the predictor $X_i, i=1,\cdots,n$ may be available only on parts of $\mathcal{T}$. We first give some notations and then make further analysis. Let $X_1,\cdots, X_n$ be independent and identically distributed samples from the random function $X$. We denote the observed and missing parts of $X_i$ by $O_i$ and $M_i$ with $O_i\cup M_i=\mathcal{T}$. Let $O_i=[L_i,R_i]\subseteq \mathcal{T}$, and assume that it is a random subinterval independent of $X_i$ with $R_i-L_i>0$ almost surely. The observed data for $i$th functional predictor is then given as  $X_i(t),t\in O_i,\,i=1,\cdots,n$, denoted by $X_{iO_i}$. In this section, our objective interest is to develop an estimation method for model \eqref{mainmodel} with partially observed functional observations without and with measurement error respectively. And in these scenarios, our objective is to get estimators of the functional principal component scores $\{U_{ij}\}$ and the eigenfunctions $\{\phi_j\}$ as indicated in formulas \eqref{gammasep}, \eqref{gammahat}. Depending on whether measurement error is presented in partially observed functional curves, two methods are developed: one is established by applying linear functionals of the observed parts of that trajectory, while the other one is based on principal component analysis through conditional expectation.

\subsection{Partially Observed Functional Data without Measurement Error} \label{wme}
In the scenario that functional curves are partially observed on the domain without measurement error, to get an estimator of $\gamma$ in model \eqref{mainmodel}, we need to get estimators of $U_{ij}$ and $\phi_j$ pertaining to this case. An estimator of $U_{ij}$ is obtained based on the linear functional of the observed part $X_{iO_i}$, and an estimator of $\phi_j$ is obtained by giving estimators of mean and covariance function of $X$. The steps are given here.

Step 1: Estimate the mean $\mu$ and the covariance function $c_X$ by sample mean and sample covariance.\par
Step 2: Estimate eigenvalues $\{\lambda_j\}$ and eigenfunctions $\{\phi_j\}$ by $\int_{\mathcal{T}} \hat{c}_X(s,t)\hat{\phi}_j(s) ds = \hat{\lambda}_j\hat\phi_j(t).$\par
Step 3: Estimate principal component scores $U_{ij}=U_{ijO_i}+U_{ijM_i}$ with $\hat U_{ijO_i} = \langle X_{iO_i} - \hat\mu_{O_i},\hat\phi_{jO_i}\rangle$, and estimate $U_{ijM_i}$ by modeling it as linear functionals of $X_{iO_i}$ given as $\hat U_{ijM_i} = \langle \hat\xi_{ijM_i},X_{iO_i} - {\hat\mu}_{O_i}\rangle$.\par
Step 4: Estimate $\gamma$ based on formulas \eqref{gammasep} and \eqref{gammahat} for $X_{iO_i}$ observed without measurement error.

We first address the problem of getting estimators of $\mu$ and $c_X$, denoted as $\hat{\mu}^{\text{NME}}$ and $\hat{c}^{\text{NME}}_X$ respectively, followed by establishing estimators of $U_{ij}$ and eigenfunctions $\phi_j$ which are denoted as $\hat{U}^{\text{NME}}_{ij}$ and $\hat{\phi}^{\text{NME}}_j$. For simplicity of presentation, we suppress the notation on ``NME" in this subsection unless otherwise stated.

Let $O_i(t) = \text{I}_{O_i}(t)$ with indicator function $\text{I}_{O_i}(t)$ being $1$ if $t\in O_i$, and 0 otherwise, and let $W_i(s,t) = O_i(s)O_i(t)$. 
The estimators of the mean function ${\mu}$ and the covariance function ${c}_X$ of $X$ obtained from the observed points $s, t$ of $X_i$,  are given by,
\begin{align*}
\hat{\mu}(t) = \frac{1}{\sum_{i=1}^n O_i(t)} \sum_{i=1}^n O_i(t) X_i(t), \tag{4} \label{mumis}
\end{align*}
\begin{align*}
\hat{c}_X(s,t) = \frac{1}{\sum_{i=1}^{n} W_i(s,t)}\sum_{i=1}^{n} W_i(s,t) (X_i(s) - \hat{\mu}(s))(X_i(t) - \hat{\mu}(t)). \tag{5} \label{covmis}
\end{align*}
Therefore, we get the estimators $\{\hat{\lambda}_j\}$, $\{\hat\phi_j\} $ related to $\{\lambda_j\}$ and $\{\phi_j\}$ from $\hat{c}_X$ associated with the covariance operator $\hat{C}_X$. 

We could not get estimators $\hat{U}_{ij}$ of FPC scores $\{U_{ij}\}$ of $X_i$ directly from its definition if $O_i\ne \mathcal{T}$. To bridge the gap,  $U_{ij}$ is decomposed into two parts:
\begin{align*}
U_{ij}=\langle X_{iO_i} - \mu_{O_i},\phi_{jO_i}\rangle+\langle X_{iM_i} - \mu_{M_i},\phi_{jM_i}\rangle=U_{ijO_i}+U_{ijM_i}, \tag{6} \label{xscore}
\end{align*}
where $\mu_{O_i}$ and $\phi_{jO_i}$ denote the restriction of $\mu$ and the eigenfunction $\phi_j$ on $O_i$ respectively, and the definitions of $\mu_{M_i}$, $\phi_{jM_i}$
are similar. The estimator $\hat{U}_{ijO_i}$ of $U_{ijO_i}$ can be estimated directly from the observed part $X_{iO_i}$ and the estimator $\hat \phi_j$, given as $\hat{U}_{ijO_i} = \langle X_{iO_i}-\hat\mu_{iO_i}, \hat \phi_{jO_i}\rangle$.
For the term $U_{ijM_i}$, we consider using the linear functional form $\langle \xi_{ijM_i},X_{iO_i} - {\mu}_{O_i}\rangle$  of the observed part $X_{iO_i}$ to estimate it which is also considered in \cite{kraus2015components}, that is, 
$$
\hat{\xi}_{ijM_i} = \underset{\xi_{ijM_i}\in L^2}{\text{argmin}}\,\, n^{-1}\sum_{i=1}^{n} (\hat{U}_{ijM_i} - \langle \xi_{ijM_i},X_{iO_i}-\hat{\mu}_{iO_i}\rangle)^2
$$
with ${\hat{U}}_{ijM_i} = \langle X_{iM_i}-{\hat{\mu}}_{M_i}, {\hat{\phi}}_{jM_i}\rangle$. The estimator ${\hat{\xi}}_{ijM_i}$ has the explicit form: $\hat{\xi}_{ijM_i}={{\hat C}_{O_iO_i}}^{-1}{\hat C}_{O_iM_i}{\hat\phi}_{jM_i}$, where $\hat C_{O_iO_i}$, $\hat C_{O_iM_i}$ are the empirical covariance operator for $C_{O_iO_i}$, $C_{O_iM_i}$ with the kernel being the covariance function $\hat c_X$ of $X_i$
restricted to $O_i\times O_i$ and $O_i\times M_i$ respectively. To obtain a stable solution, we adopt ridge regularization, given by  
\begin{align*}
&\hat\xi_{ijM_i}^{(\rho)}={(\hat C^{(\rho)}_{O_iO_i})}^{-1}\hat C_{O_iM_i}\hat\phi_{jM_i},\\
&\hat U^{(\rho)}_{ijM_i}=\langle\hat\xi_{ijM_i}^{(\rho)},X_{iO_i} - \hat{\mu}_{iO_i}\rangle,
i=1,\cdots,n,\,j=1,\cdots,m,  \tag{7} \label{scorehat}
\end{align*}
where $\hat{C}_{O_iO_i}^{(\rho)}=\hat{C}_{O_iO_i}+\rho\mathcal{F}_{O_i}$, $\mathcal{F}_{O_i}$ is an identity operator defined on $L^2(O_i)$, $\rho$ is a ridge parameter; see \cite{kraus2015components} for further details. Let $\hat{U}^{\text{NME}}_{ij} = \hat{U}_{ijO_i} + \hat U^{(\rho)}_{ijM_i}$.
The estimator $\hat{\gamma}^{\text{NME}}$ of $\gamma$ using all of the information of the dataset is then obtained through replacing $\hat{U}_{ij}$ in \eqref{gammasep} with $\hat{U}^{\text{NME}}_{ij}$,
\begin{align*}
\hat {\gamma}^{\textrm{NME}}(t)=\sum_{j=1}^m \hat{\gamma}_j\hat\phi_{j}(t).  \tag{8} \label{gamma2}
\end{align*}

To facilitate our theoretical analysis, we first impose some assumptions on observation points for partially observed functional curves, indicating the observation points asymptotically provide enough information in individual or pairwise crossover.

(A1) $\text{There~ exists} ~\delta_1>0  ~\text{s.t.} \mathop{\text{sup}}\limits_{t\in[0,1]}\text P\{n^{-1}\sum_{i=1}^n \text{I}_{O_i}(t)\leq\delta_1\}=O(n^{-2})$.\par 
(A2) $\text{There~ exists} ~\delta_2>0 ~\text{s.t.} \mathop{\text {sup}}\limits_{s,t\in[0,1]^2}\text P\{n^{-1}\sum_{i=1}^nW_i(s,t)\leq\delta_2\}=O(n^{-2}). $\par
Moreover, we also introduce some regularity conditions necessary to derive theoretical properties for the estimate $\hat {\gamma}^{\text{NME}}$. 

(A3) $\text E||X-\mu||^4<\infty$.\par \label{A}
(A4) $nm^{-1}\rightarrow\infty$, $n/(\sum_{j=1}^{m}\delta^{-2}_j)\rightarrow\infty$ with $\delta_j=\text{min}_{ j\geq 1}\{\lambda_{j}-\lambda_{j+1}, \lambda_{j-1}-\lambda_{j} \}$ and $n\lambda^2_m\rightarrow\infty$ as $m\rightarrow\infty$.\par 
(A5) The ridge parameter $\rho$ satisfies $ \rho\rightarrow 0$, $n\rho^3\rightarrow 0$, $nm^{-1}\rho^2\rightarrow\infty$.\par
(A6) $\sum_{k=1}^{\infty}[\text{E}[YU_k]]^2/\lambda_k^2<\infty$. \par
(A7) $\sum_{j=1}^\infty\sum_{k=1}^\infty\frac{r_{M_iO_ijk}^2}{\lambda_{O_iO_ik}^2}<\infty,$ with $r_{M_iO_ijk}=\text{cov}(\langle X_{M_i}-\mu_{M_i},\phi_{M_iM_ij}\rangle, \langle X_{M_i}-\mu_{M_i}, \phi_{O_iO_ik}\rangle)$. \par

Assumption (A3) is a common condition in the analysis of functional model by using the method of FPCA to guarantee the random functions have finite fourth moment (see \cite{cardot1999functional}). Note that if the eigenvalues $\{\lambda_j\}$ are exponentially or geometrically decreasing, the assumption (A4) holds. The same kind of conditions are also introduced in \cite{cardot1999functional}. Assumption (A5) is used to control the size of ridge effect. To define the convergence of the right hand of the formula $\gamma(s)=\sum_{k=1}^{\infty}(\text{E}[YU_k]/\lambda_k) \phi_k(s)$, in the $L^2$ sense, assumption (A6) is required that is similar to the condition (A1) in \cite{yao2005bfunctional}.  Assumption (A7) is used to make the solution $\hat{\xi}_{ijM_i}$ valid which is commonplace in the theory of inverse problems as Picard condition (see \cite{hansen1990discrete}). 

Let $\theta_n=\sum_{k=m}^{\infty}[\text{E}[YU_k]]^2/\lambda_k^2$. Then assumption (A6) indicates that $\theta_n\rightarrow 0$. Denote $\upsilon= \sum_{j=1}^{m}V_{ij}$ with $ V_{ij}= \langle\phi_{jM_i}, (C_{M_iM_i}- C_{M_iO_i}C^{-1}_{O_iO_i}C_{O_iM_i})\phi_{jM_i}\rangle. $ Based on the above assumptions,  Theorem \ref{thm1} gives the converge rate for the estimator $\hat{\gamma}^{\text{NME}}$ in the $L^2$ sense.
\bo \label{thm1}
	Suppose that (A1)-(A7) are satisfied. Then 
	\begin{align*}
	\|\hat{\gamma}^{NME}-\gamma\|^2 = O_p(n^{-1}m\rho^{-2} + \iota_n + \theta_n + \upsilon).
	\end{align*}
	with $\iota_n = n^{-1}\sum_{j=1}^{m} \delta^{-2}_j$.
\eo

Theorem \ref{thm1} indicates that the approximation error rate of $\hat{\gamma}^{\text{NME}}$ for $\gamma$ is controlled by four terms. The first term depends on sample size $n$, tuning parameter $m$, ridge parameter $\rho$, which is of the higher order than the one given in \cite{hall2007methodology} that is mainly due to functional curves observed on the part of the domain. The second term is related to the spacings between adjacent eigenvalues, and its effect on convergence rate of $\gamma$ is also emphasized in \cite{hall2007methodology}.
The third term is related to the convergence of $\gamma$ in $L^2$ sense, which is also show in \cite{yao2005bfunctional} to get approximation error rate for functional coefficient. The fourth term is introduced by approximating $U_{ijM_i}$ with $\tilde{U}_{ijM_i}$. 

Note that in practice, the ridge parameter $\rho$ included in the regularized estimation of the $j$th score of the $i$th functional observation is chosen by generalized cross-validation based on the set of samples observed on the entire domain (see \cite{kraus2015components}).

\subsection{Partially Observed Functional Data  with Measurement Error}\label{mme}
In this subsection, we construct an estimator for the slope function $\gamma$ for partially observed trajectories with measurement error. We suppose the functional observations are: 
\begin{align*}
Z_{il} = X_i(t_{il})+\varepsilon_{il},\quad t_{il}\in O_i, i=1,\cdots,n, l=1,\cdots N_i, \tag{9} \label{obme}
\end{align*}
where $\varepsilon_{il}$ is independent from all the other variables $X_j, j\neq i$, with $\text{E}(\varepsilon_{il}) = 0$, $\text{var}(\varepsilon_{il})=\sigma^2_X$.

To get an estimator of $\gamma$ in \eqref{mainmodel} in the scenario that trajectories may be observed on parts of the domain with measurement error (WME), we need give estimators of FPC scores and eigenstructure pertaining to this case. Estimator of eigenstructure is established after using local linear smoothers to get estimators of mean and covariance function of $X$. We obtain estimators of FPC scores by using approach of principal component analysis via conditional expectation. The steps are given here.

Step 1: Estimate the mean and covariance functions by local linear smoothers.\par
Step 2: Estimate eigenvalues $\{\lambda_j\}$ and eigenfunctions $\{\phi_j\}$ by\\ $\int_{\mathcal{T}} \hat{c}^{\text{WME}}_X(s,t)\hat{\phi}^{\text{WME}}_j(s) ds = \hat{\lambda}^{\text{WME}}_j{\hat\phi}^{\text{WME}}_j(t)$.\par
Step 3: Estimate FPC scores $\{U_{ij}\}$ by principal component analysis via conditional expectation (PACE): $\tilde{U}_{ij} = \text{E}[U_{ij}|\textbf{Z}_i]$.\par
Step 4: Based on obtained estimators $\hat{\tilde{U}}_{ij}$ and $\hat{\phi}^{\text{WME}}_j$, we get estimator $\gamma^{\text{WME}}$ for $X_{iO_i}$ observed with measurement error.

We first calculate estimators for the mean and the covariance function of $X$ in the scenario \eqref{obme}, denoted as $\hat{\mu}^{\text{WME}}$ and $\hat{c}^{\text{WME}}_X$, that are required to derive estimators for the FPC scores $U_{ij} = \int (X_i(t)-\mu(t))\phi_j(t)\text{dt}$. For simplicity of presentation, we suppress notation on ``WME" unless otherwise stated in this subsection. 

Let $K(\cdot)$ be a nonnegative univariate kernel function that is assumed to be a symmetric probability density function (pdf) with compact support $\text{supp}(K)=[-1,1]$, and $h_{\mu}$, $h_c$ be the bandwidths for obtaining estimators of ${\mu}$, ${c}_X$. Assume that the second derivatives of $\mu$, $c_X$ on $\mathcal{T}$, $\mathcal{T}^2$ respectively exist. We use local linear smoothers for the mean function $\mu$ (\cite{yao2005afunctional}, \cite{yao2005bfunctional}, \cite{kneip2020optimal}) defined as $\hat{\mu}(t) = \hat{\beta}_0$, where 
\begin{align*}
(\hat{\beta}_0,\hat{\beta}_1) = \underset{\beta_0,\beta_1}{\text{argmin}}\sum_{i=1}^{n}\sum_{l=1}^{N_i}K\left(\frac{t_{il}-t}{h_{\mu}}\right)[Z_{il}-\beta_0-\beta_1(t-t_{il})]^2. \tag{10} \label{meanest}
\end{align*}
Let $\hat G_{ilk} = (Z_{il}-\hat{\mu}(t_{il}))(Z_{ik}-\hat{\mu}(t_{ik}))$ be the raw covariance points. 
The local linear smoother for the covariance function $c_X$ is defined as $\hat{c}_X = \hat{\tilde{\beta}}_{0}$, where 
\begin{align*}
(\hat{\tilde{\beta}}_{0},\hat{\tilde{\beta}}_{1},\hat{\tilde{\beta}}_{2}) = \underset{\tilde{\beta}_{0},\tilde{\beta}_{1},\tilde{\beta}_{2}}{\text{arg\,min}}\,\sum_{i=1}^{n}\sum_{1\leq l,k\leq N_i}
& K\left(\frac{t_{il}-t}{h_c}\right)K\left(\frac{t_{ik}-s}{h_c}\right) \\ 
&\times [\hat{G}_{ilk}-\tilde{\beta}_{0}-\tilde{\beta}_{1}(t_{il}-t)-\tilde{\beta}_{2}(t_{ik}-s)]^2. \tag{11} \label{covest}
\end{align*}
Similar to the technique introduced in \cite{yao2005afunctional}, the points $\hat G_{ill}, l=1\cdots,N_i$ are not included in \eqref{covest}. Let $\mathcal{T}_1=[\text{inf}\{L_i \in \mathcal{T}, i=1,\cdots,n\}+|\mathcal{T}|/4, \text{sup}\{R_i\in\mathcal{T}, i=1,\cdots,n\}-|\mathcal{T}|/4]$ with  $|\mathcal{T}|$ being the length of $\mathcal{T}$. The estimator of $\sigma^2_X$ is defined as $\hat{\sigma}^2_X$ if $\hat{\sigma}^2_X>0$, otherwise $\hat{\sigma}^2_X=0$ with $$\hat{\sigma}^2_X=2\int_{\mathcal{T}_1}(\hat{V}_X(t)-\tilde{G}(t))\text{dt}/|\mathcal{T}|,$$ where $\hat{V}_X(t)$ is the local linear estimator using the points $\{\hat G_{ill}\}$,  $\tilde{G}(t)$ is the estimate $\hat{c}_X(s,t)$ restricted to $s=t$ (\cite{staniswalis1998nonparametric}, \cite{yao2005afunctional}). The estimators of $\{\lambda_j, \phi_j\}_{j\geq1}$ are the corresponding solutions of the eigen-equations 
\begin{align*}
\int_\mathcal{T} \hat{c}_X(s,t)\hat{\phi}_j(s) \text{ds} = \hat{\lambda}_j\hat{\phi}_j(t).
\end{align*}

Based on the K-L expansion of $X_i$, model \eqref{obme} can be rewritten as 
\begin{align*}
Z_{il} = \mu(t_{il}) + \sum_{j=1}^{\infty}U_{ij}\phi_{j}(t_{il}) + \varepsilon_{il}, \quad t_{il}\in O_i, i=1\cdots,n, l=1\cdots,N_i.
\end{align*}
Let ${\mathbf{X}}_i=(X_i(t_{i1}),\cdots,X_i(t_{iN_i}))^T$, ${\mathbf{Z}}_i=(Z_{i1},\cdots,Z_{iN_i})^T$, ${\bm{\mu}}_i=(\mu(t_{i1}),\cdots,\mu(t_{iN_i}))^T$, ${\bm{\phi}}_{ij}=(\phi_j(t_{i1}),\cdots,\phi_j(t_{iN_i}))^T$. Assume that $U_{ij}$ and $\varepsilon_{il}$ are jointly Gaussian. Following \cite{yao2005afunctional}, the best prediction of $U_{ij}$ of the $i$th subject given the observations ${(Z_{il}, t_{il}), l=1,\cdots,N_i}$ is obtained as
\begin{align*}
\tilde{U}_{ij} = \lambda_j\bm\phi^T_{ij}\bm\Sigma^{-1}_{{\mathbf{Z}}_i}({\mathbf{Z}}_i-\bm{\mu}_i),
\end{align*}
where $\bm\Sigma_{{\mathbf{Z}}_i}=\text{cov}({\mathbf{Z}}_i, {\mathbf{Z}}_i) = \text{cov}({\mathbf{X}}_i, {\mathbf{X}}_i) + \sigma^2_X \mathbf{I}_{N_i}$ with identity matrix $\mathbf{I}_{N_i}$. That is, the $(u,v)$th element of $\bm\Sigma_{{\mathbf{Z}}_i}$ is $(\bm\Sigma_{{\mathbf{Z}}_i})_{u,v} = c_X(t_{iu}, t_{iv}) + \sigma^2_XI_{uv}$ with $I_{uv}=1$ if $u=v$, and 0 otherwise. Then the estimator of ${U}_{ij}$ is given through substituting $\mu, \lambda_j,\phi_j$ with $\hat{\mu}, \hat{\lambda}_j, \hat{\phi}_j$ as 
\begin{align*}
\hat{U}^{\textrm{WME}}_{ij} = \hat\lambda_j\hat{\bm\phi}^T_{ij}\hat{\bm\Sigma}^{-1}_{{\mathbf{Z}}_i}({\mathbf{Z}}_i-\hat{\bm{\mu}}_i), \tag{12} \label{pace}
\end{align*}
where the $(u,v)$th entry of $\hat{\bm\Sigma}_{{\mathbf{Z}}_i}$ is  $(\hat{\bm\Sigma}_{{\mathbf{Z}}_i})_{u,v} = \hat{c}_X(t_{iu}, t_{iv}) + \hat{\sigma}^2_X I_{uv}$.
Replacing $\hat{U}_{ij}$ in \eqref{gammasep} with $\hat{U}^{\text{WME}}_{ij}$, we then get the estimator $\hat{\gamma}^{\text{WME}}$ of $\gamma$ from \eqref{gammahat}
\begin{align*}
\hat{\gamma}^{\textrm{WME}}(t) = \sum_{j=1}^{m} \hat{\gamma}_j \hat{\phi}_j,
\end{align*} 
where $\hat{\gamma}_j$ is the $j$th entry of $\hat{\bm\gamma}$ with $\hat{U}^{\text{WME}}_{ij}$ in \eqref{gammasep}.

Next, we give some theoretical results for $\hat{\gamma}^{\text{WME}}(t)$. We assume the following regularity conditions which are similar to the assumptions in \cite{kneip2020optimal}, \cite{yao2005bfunctional}.

(B1) The observational points $\{t_{il}, l=1,\cdots,N_i\}$ given $O_i$ for the $i$th subject, are i.i.d. random variables with pdf $f_{t|O_i}(u)>0$ for all $u\in O_i\subseteq \mathcal{T}$ and zero else. For the marginal pdf $f_t$ of observation times $t_{ij}$, $f_t(u)>0$ for all $u\in\mathcal{T}$.\par 
(B2) Let $N=\text{min}\{N_i, i=1,\cdots, n\}$. $N\asymp n^r$ with $0<r<\infty$, where $a_n\asymp b_n$ means that there exists a constant $0<L<\infty$ such that $a_n/b_n\rightarrow L$ as $n\rightarrow\infty$.\par
(B3) $h_{\mu}\rightarrow 0$, $h_c\rightarrow 0$, $nNh_{\mu}\rightarrow\infty$, $nMh_c\rightarrow\infty$ as $n\rightarrow\infty$ with $M=N^2-N$.\par
(B4) $K$ is a second order kernel with compact support $[-1,1]$.\par
(B5) Let $G_{ilk} = (Z_{il}-{\mu}(t_{il}))(Z_{ik}-{\mu}(t_{ik}))$. Define $f_{Zt}$, $f_{tt}$, $f_{Gtt}$ as the joint pdf of $(Z_{il}, t_{il})$ on $\mathbb{R}\times\mathcal{T}$, $(t_{il_1},t_{il_2})$ on $\mathcal{T}^2$, $(G_{ilk}, t_{il},t_{ik})$ on $\mathbb{R}\times \mathcal{T}^2$, respectively. All of the second derivatives of $f_{Zt}$, $f_{tt}$, $f_{Gtt}$ are uniformly continuous and bounded. Moreover, $f_{t}$ is uniformly  continuous and bounded on $\mathcal{T}$.\par
(B6) Let $\mathbf{\Lambda}=\text{diag}\{\lambda_1,\cdots,\lambda_m\}$, $\Xi =(\lambda_1\mathbf{{\bm{\phi}}}_{i1},\cdots,\lambda_m\mathbf{{\bm{\phi}}}_{im})^T$, $\Upsilon = \mathbf{\Lambda}-\Xi\mathbf{\Sigma}^{-1}_{{\mathbf{Z}}_i}\Xi^T$ and $\varsigma_n\equiv \text{trace}(\Upsilon)$. 
Denote $r_{\mu}=h^2_{\mu} + 1/\sqrt{nNh_{\mu}} + 1/\sqrt{n}$, $r_{c} = h^2_{c} + 1/\sqrt{nMh^2_{c}} + 1/\sqrt{n}$. $\upsilon_n \equiv mr_{\mu}\rightarrow 0$,  $\tau_n \equiv r_c(\sum_{j=1}^{m}\delta^{-1}_j)\rightarrow 0$.

\bo \label{thm2}
	Under the regularity conditions (A3), (A6), (B1)-(B6), we have that 
	\begin{align*}
	\|\hat{\gamma}^{\mathrm{WME}}-\gamma\|^2 = O_p(\upsilon_n+\tau_n+\varsigma_n+\theta_n).
	\end{align*}
\eo

Theorem \ref{thm2} gives the rate of convergence of the estimator $\hat{\gamma}^{\text{WME}}$ in the $L^2$ sense. The rate of convergence of $\hat{\gamma}^{\text{WME}}$ depends on the sample size and bandwidths which is common for estimating curves or surface by local linear smoothers for functional data analysis (see \cite{li2010uniform}). Related results of Theorem \ref{thm2} can also be found in \cite{yao2005bfunctional}. The terms $\upsilon_n, $ $\tau_n$ are related to rates of convergence of estimators for the mean and covariance function by using local linear smoothers. The term $\varsigma_n$ are introduced by approximating $U_{ij}$ with $\tilde{U}_{ij}$.

\section{Simulation Studies}\label{simu}
In this section, we use the simulated datasets to evaluate the finite sample properties of our proposed methods in Section \ref{method}. This studies are based on $n\in\{50,100,200\}$ i.i.d. samples $\{X_i,Y_i\}_{i=1}^n$ and equally spaced grid $\{t_1,\cdots,t_{30}\}$ on $[0,1]$ with $t_1=0, t_{30}=1$. For the $i$th functional observation $X_i(t)$, the missing interval $M_i$ takes the form $[R_i-E_i,R_i+E_i]$, with $R_i=a_1T_{i1}^{1/2}$, $E_i=a_2T_{i2}$, where $T_{i1},\,T_{i2}$ are independent random variables uniformly distributed on $[0,1]$, $a_1,a_2\in\mathbb{R}$. We consider $(a_1,a_2)=(1.5,0.2)$, $(a_1,a_2)=(1.5,0.4)$ with the expected missing length over the domain being $0.4$ and $0.8$, respectively. We set the intercept $\alpha=0$. To evaluate the performance of an estimator $\hat{\gamma}$ of $\gamma$, mean integrated square error (MISE) is used below as an evaluation criterion, given by, $$\text{MISE} = \frac{1}{N}\sum_{l=1}^{N}\int_{0}^{1}(\hat\gamma_l(t)-\gamma(t))^2\text{dt},$$ where $N$ is the number of Monte Carlo replications.

For functional predictors $\{X_i\}$ without measurement error, the trajectories are generated as follows. The simulated random function $X_i$ has zero mean, the covariance function is generated from two eigenfunctions, $\phi_1(t)=\sqrt{2}\text{sin}(\pi t/2)$, $\phi_2(t)=\sqrt{2}\text{sin}(3\pi t/2)$. For the eigenvalues, we take $\lambda_1 = (\pi/2)^{-2}$, $\lambda_2=(3\pi/2)^{-2}$, $\lambda_k=0,$ for $k\geq 3$. The error $\epsilon_i$ in \eqref{mainmodel} is assumed to be standard normal. For the slope function $\gamma$ in \eqref{mainmodel}, we take the form $\gamma(t)=\phi_1(t)+3\phi_2(t)$. We compare the finite sample performance of our proposed method with the method that gives an estimator for $\gamma$ through formula \eqref{gammasep}, \eqref{gammahat} with deleting the incomplete functional observations in the datasets and denote it as ``SUB". Moreover, the estimator of $\gamma$ based on the original complete dataset is also considered in this scenario, and denote it as ``ORI". We conduct $1000$ simulation runs in each setup. Table \ref{tab1} reports the results.

\begin{table}[htp!]
	\caption{\textmd{MISEs of the estimates of $\gamma$ under different methods with 1000 Monte Carlo replications for functional predictors without measurement errors.}}
	\begin{center} \label{tab1}
		\renewcommand\tabcolsep{8.0pt} 
		\renewcommand\arraystretch{1.5} 
		\begin{threeparttable}
			\begin{tabular}{cccccccccccc}\hline
				method  &$(a_1, a_2)$&  {$n=50$} &  {$n=100$} & {$n=200$}\\
				\hline
				{ORI \tnote{1}}&  &2.0295  &1.0767  &0.3670 \\
				\hline
				{NME \tnote{2}}& (1.5, 0.2) &2.8653  &1.6650 &0.7343\\
				& (1.5, 0.4) &3.5650  &2.4412 &1.3497 \\
				\hline
				{SUB} \tnote{3} & (1.5, 0.2) &3.5632  &1.8844 &0.8322 \\
				& (1.5, 0.4) &4.600  &2.6664 &1.4401 \\
				\hline
			\end{tabular}
			\begin{tablenotes}
				\item [1] The estimator is obtained with the original dataset $\{X_i,Y_i\}$ with functional predictors observed in entire domain $[0,1]$ (ORI).
				\item [2] The estimator $\hat{\gamma}^{NME}$ introduced in section \ref{wme} (NME).
				\item [3] The estimator is obtained by deleting the functional predictors with missing parts (SUB).
			\end{tablenotes}
		\end{threeparttable}
	\end{center}
\end{table}

As shown in Table \ref{tab1}, in the scenario where incomplete functional predictors are observed without measurement error, the estimation method in Section \ref{wme} performs better than ``SUB" method. This is because some useful information the dataset has will be lost if we delete them directly, while the ``NME" method can take advantage of the whole information about the dataset. Specially, in each setting for $(a_1,a_2)$, MISEs from the ``NME" method have smaller values relative to the ``SUB" method. These simulation results also demonstrate that MISEs decrease with increasing sample size $n$ for these three methods. And MISEs increase with longer missing length on $[0,1]$ at fixed $n$ indicating that a large error is  introduced for the ``NME" method in imputing missing scores of incomplete functional predictors through little available information from functional samples.  In further, the difference of MISEs among these three methods are reduced with increasing sample size $n$, and the ``NME" method still performs better than the ``SUB" method, those imply the ``NME" method is promising. 

For functional predictors $X_i$ with measurement error, they are generated according to $Z_i(t_{il}) = X_i(t_{il})+\varepsilon_{il}, l=1,\cdots,30$, as follows. We take $X_i(t)=\sum_{j=1}^{50}U_{ij}\phi_j(t)$ with $U_{ij}=(-1)^{j+1}j^{-1.1/2} W_{ij}$, where $W_{ij}$ is uniformly distributed on $[-\sqrt{3}, \sqrt{3}]$, $\phi_1(t)=1$, $\phi_j(t)=\sqrt{2}\text{cos}(j\pi t)$ for $j\geq 2$. The additional random error $\varepsilon_{il}, l=1\cdots,30$ and the error $\epsilon_i$ in \eqref{mainmodel} are assumed to be normal with mean zero, variance $0.25$. For the slope function $\gamma$, we take $\gamma=\sum_{j=1}^{50}\gamma_j\phi_j(t)$ with $\gamma_1=0.3$, $\gamma_j=4(-1)^{j+1}j^{-2}$ for $j\geq 2$ (\cite{hall2007methodology}). We conduct $100$ simulation runs in each setup. To demonstrate the superior performance of our proposed method in Section \ref{mme}, we compare it with the other two methods after we get estimators of $\mu(t)$ and $c_X(s,t)$ by solving the optimization problems \eqref{meanest}, \eqref{covest} respectively: one is that an estimator of $\gamma$ is established by applying integration method to get the FPC scores $\hat{U}_{ij}$ in \eqref{gammasep} instead of using formula \eqref{pace}, denoted as ``IN"; the other one is that an estimator of $\gamma$ is obtained by using the method in Section \ref{wme} with dataset $\{Z_i, Y_i\}$ with measurement error being ignored. The results are summarized in Table \ref{tab2}.

\begin{table}[htp!]
	\caption{\textmd{MISEs of the estimates of $\gamma$ under different methods with 100 Monte Carlo replications for functional predictors with measurement errors.}}
	\begin{center} \label{tab2}
		\renewcommand\tabcolsep{8.0pt} 
		\renewcommand\arraystretch{1.5} 
		\begin{threeparttable}
			\begin{tabular}{cccccccccccc}\hline
				method  &$(a_1, a_2)$&  {$n=50$} &  {$n=100$} & {$n=200$}\\
				\hline
				{WME \tnote{1}}
				&(1.5, 0.2) &0.1535 &0.1176  &0.0753 \\
				&(1.5, 0.4) &0.2033 &0.1607  &0.1057 \\
				\hline
				{IN \tnote{2}}
				& (1.5, 0.2) &0.1702  &0.1560 &0.1024 \\
				& (1.5, 0.4) &0.2671  &0.2374 &0.1974 \\
				\hline
				{NME} \tnote{3} 
				& (1.5, 0.2) &0.6312 &0.4517 &0.3320 \\
				& (1.5, 0.4) &0.7249 &0.5086 &0.3808 \\
				\hline
			\end{tabular}
			\begin{tablenotes}
				\item [1] The estimator is obtained by using the method in section \ref{mme} (CE).
				\item [2] The estimator is obtained by using integration method to get estimators of the principal component scores ${U}_{ij}$ (IN).
				\item [3] The estimator is obtained by using the method in section \ref{wme} (NME).
			\end{tablenotes}
		\end{threeparttable}
	\end{center}
\end{table}

We find from Table \ref{tab2} that the ``WME" method has the best performance relative to the other two methods in each setup, and the gains are dramatic when switching from the ``NME" method to the ``WME" method with the ``NME" method ignoring observation errors for functional predictors. Specifically, for the case of $n=100$, comparing with the ``NME" method,  the MISEs are reduced by 74\%, 68\%  using the ``WME" method with $(a_1,a_2)=(1.5,0.2)$ and $(a_1,a_2)=(1.5,0.4)$ respectively. For the ``IN" method, it provides an reasonable estimator for $\gamma$ and has better performance than the ``NME" method, but nevertheless the ``WME" method still performs better than ``IN" method with improvement of 25\%, 32\% with respect to $(a_1,a_2)=(1.5,0.2)$ and $(a_1,a_2)=(1.5,0.4)$. In addition, these simulation results show that the MISEs decrease with increasing sample size $n$ that is consistent with the derived theoretical results.

To sum up, in the scenario that incomplete functional predictors observed without measurement error, the ``NME" method taking advantage of the whole information of the dataset produces a better estimator compared with the ``SUB" method; in the scenario that incomplete functional predictors observed with measurement error, the ``WME" method is preferred for giving the smallest MISE relative to the ``IN" and ``NME" methods. Both MISEs of the estimators of $\gamma$ decrease with increasing sample size $n$, that is consistent with the derived theoretical properties.

\section{Real Data Analysis}
A real diffusion tensor imaging (DTI) dataset considered here is from NIH Alzheimer's Disease Neuroimaging Initiative (ADNI) study with 212 subjects, and is obtained through http://adni.loni. usc.edu/. The primary goal of ADNI study is to test whether serial magnetic resonance imaging (MRI), positron emission tomography (PET), biological markers, and neuropsychological assessment can be combined to measure the progression of mild cognitive impairment (MCI) and early Alzheimer's disease (AD). DTI obtained using mathematical method to represent the anisotropic diffusion of the water molecule in brain organization, can be used to learn MCI and AD. The concrete measure of anisotropy include fractional anisotropy (FA), relative anisotropy (RA), Volume ratio (VR), and FA is commonly adopted for its advantage in contrast ratio of grey-white matter. More details about preprocessing and methods of this study can be found in \cite{zhu2012multivariate}, \cite{yu2016partial}. 

Our main interest is characterizing the dynamic relationship between FA and mini-mental state examination (MMSE) score which is seen as a reliable and valid clinical measure in quantitatively assessing the severity of cognitive impairment. FA is measured at 83 equally spaced grid along the corpus callosum (CC) fiber tract that is the largest fiber tract in human brain, and is responsible for much of the communication between two hemispheres, and connects homologous areas in two cerebral hemispheres.

\begin{figure}[htp!] 
	\begin{minipage}[t]{0.45\linewidth}
		\centering
		\includegraphics[height=4.0cm,width=5.0cm]{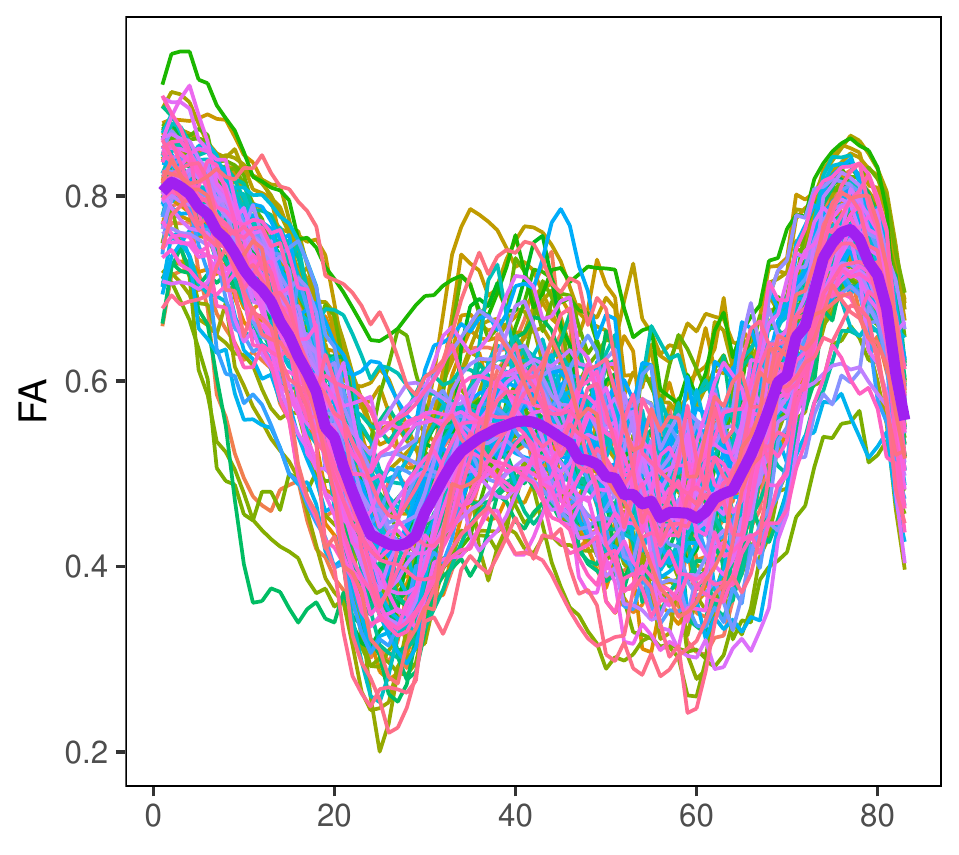}
	\end{minipage}%
	\hfill
	\begin{minipage}[t]{0.45\linewidth}
		\centering
		\includegraphics[height=4.2cm,width=5.0cm]{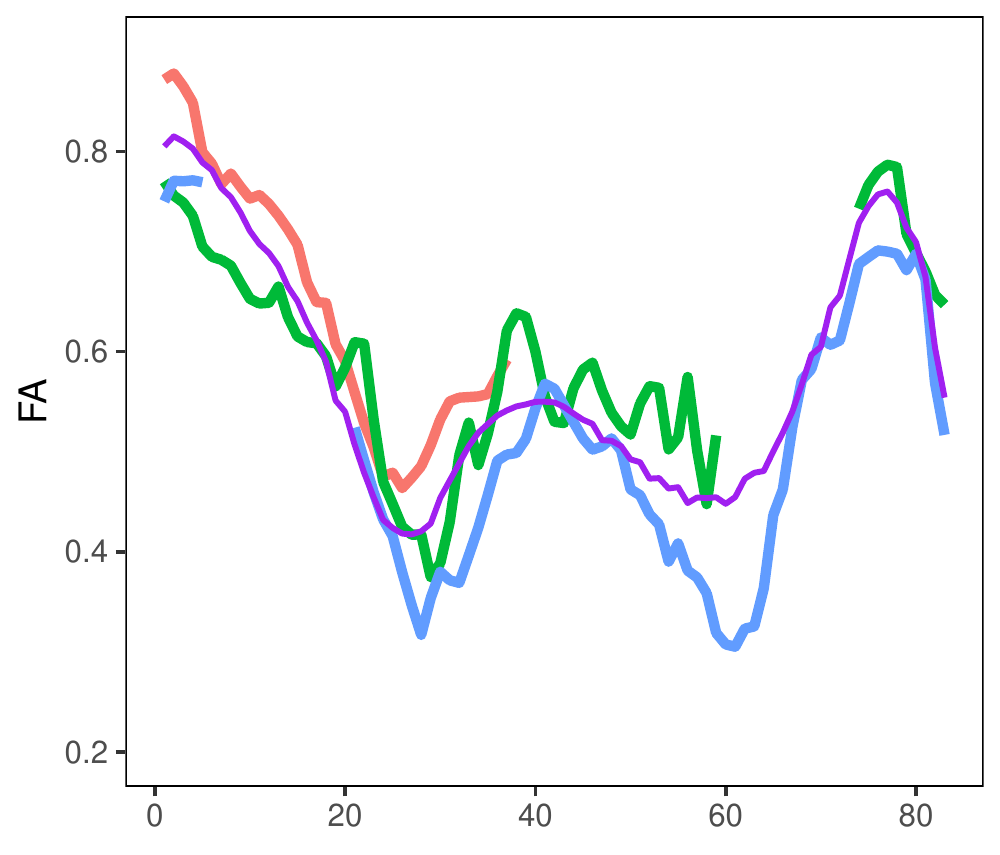}
	\end{minipage}
	\caption{\textmd{A part of complete (left) and incomplete (right) FA curves with mean function (purple line)}}
	\label{faori}
\end{figure}

To demonstrate the usefulness of the proposed method in Section \ref{wme}, we artificially delete some observed points of FA, and then compare the estimator of $\gamma$ obtained by using these incomplete functional observations with the estimator obtained by applying original complete dataset. For the $i$th FA curve, the missing domain has the same form with the interval given in Section \ref{simu} with $(a_1,a_2)=(1.5,0.2)$ and $(a_1,a_2)=(1.5,0.4)$. A part of complete and incomplete individual trajectories are displayed in Figure \ref{faori}.

\begin{figure}[htp!]
	\centering
	\includegraphics[height=5cm,width=7cm]{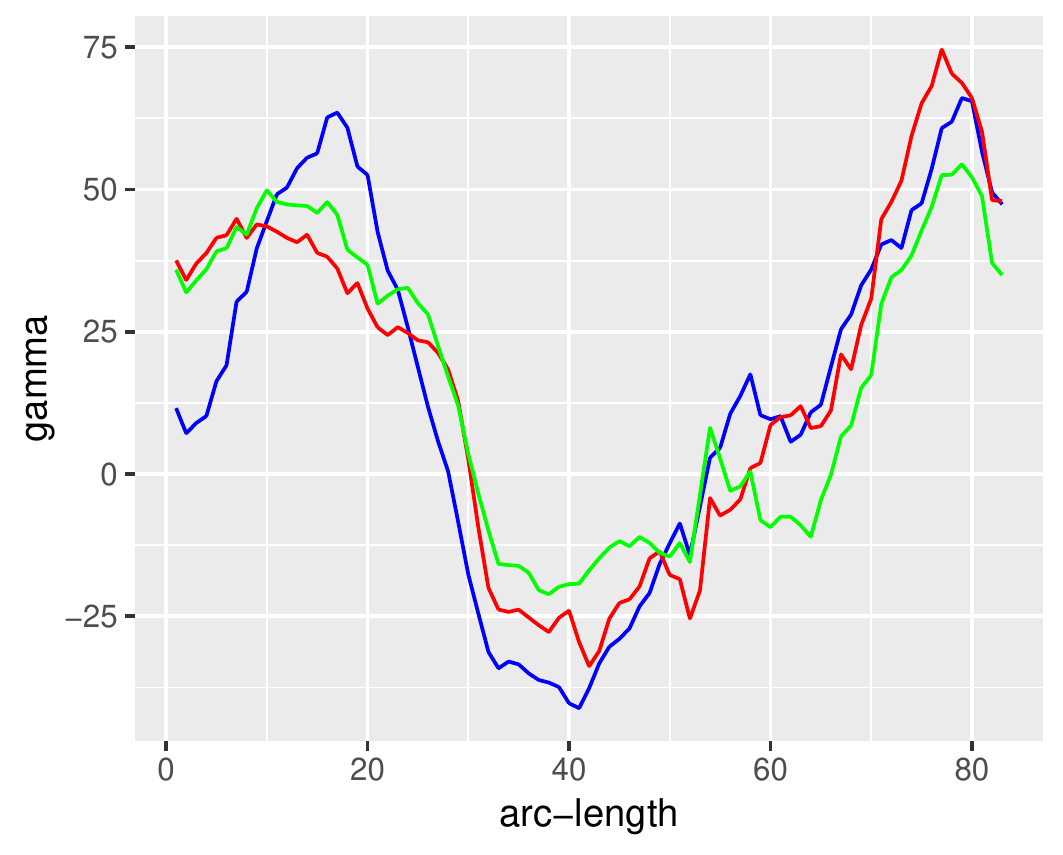}
	\caption{\textmd{Estimators of $\gamma$ with different expected missing length on $[0,1]$.  Blue line: the estimator using original complete dataset; Red line: the estimator with $(a_1,a_2) = (1.5, 0.2)$; Green line: the estimator with $(a_1,a_2) = (1.5, 0.4)$}}
	\label{fagamma}
\end{figure}

Estimators of functional coefficient obtained by both complete and incomplete FA dataset are illustrated in Figure \ref{fagamma}. It shows that estimators obtained by incomplete dataset with different missing domain (red line and green line) are similar to the estimator obtained from original complete dataset (blue line). This reveals that the proposed framework is useful in getting an estimator for the model with incomplete functional predictors.

\begin{figure*}[htp!]
	\centering
	$\begin{array}{rl}
	\includegraphics[height=4.5cm,width=5.5cm]{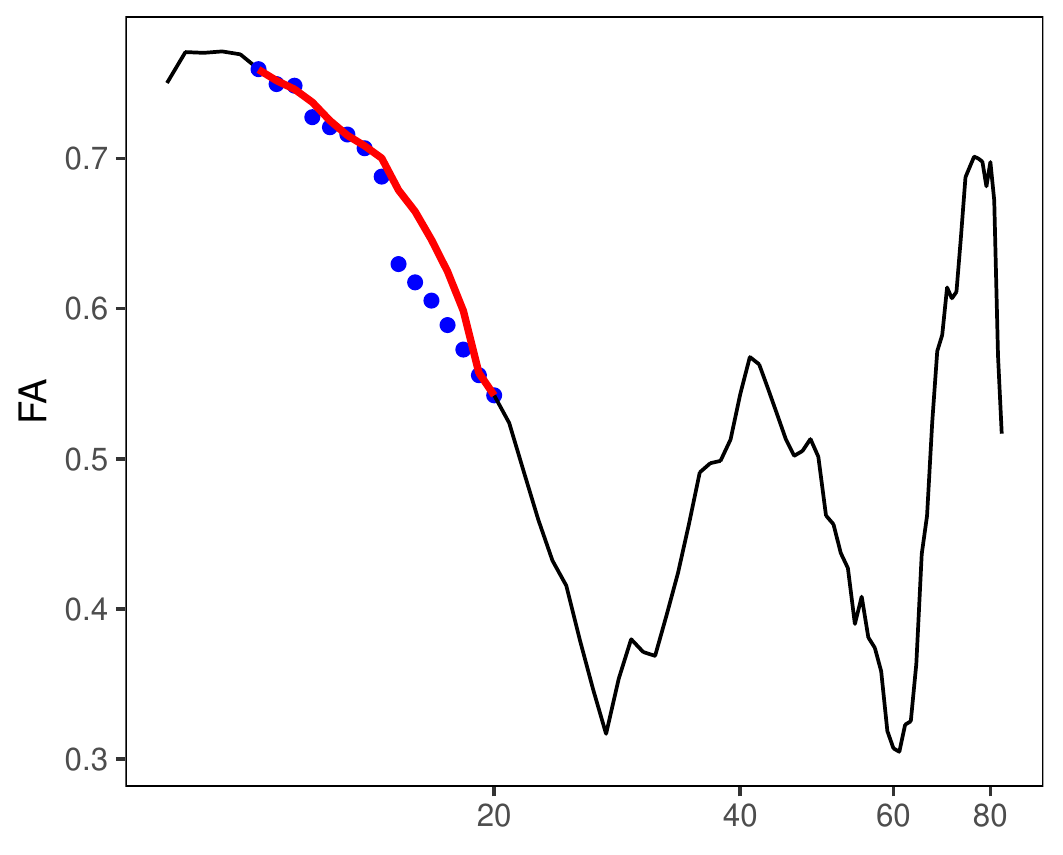} & \hspace*{0.7cm}
	\includegraphics[height=4.5cm,width=5.5cm]{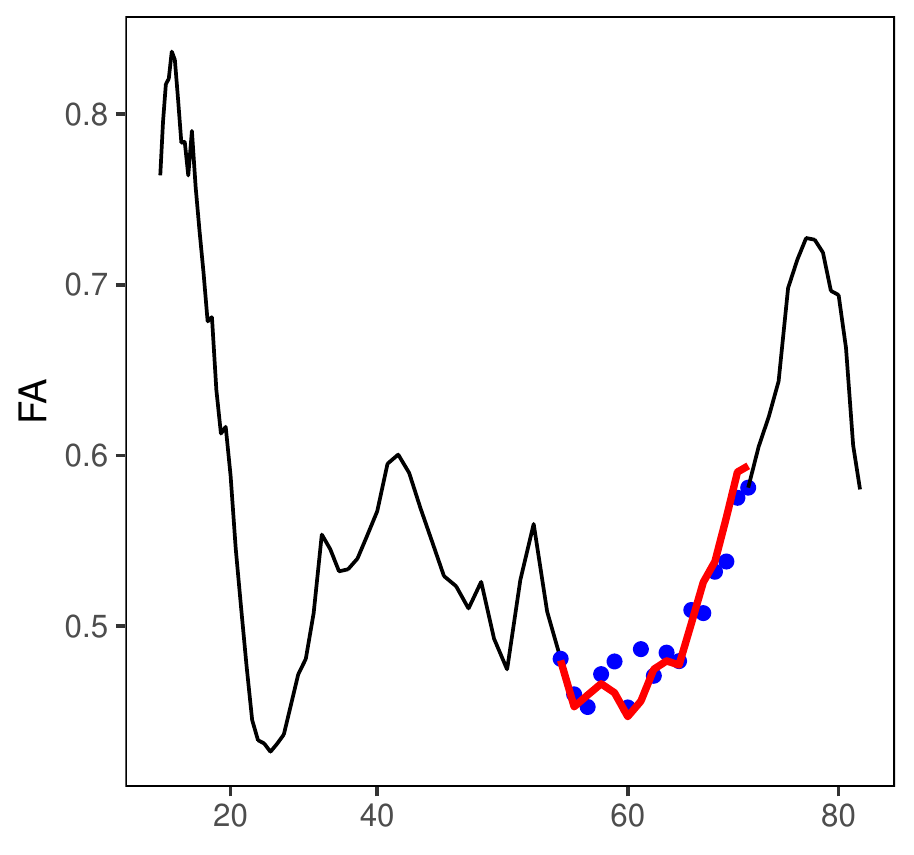} \vspace*{1cm} \\ 
	\includegraphics[height=4.5cm,width=5.5cm]{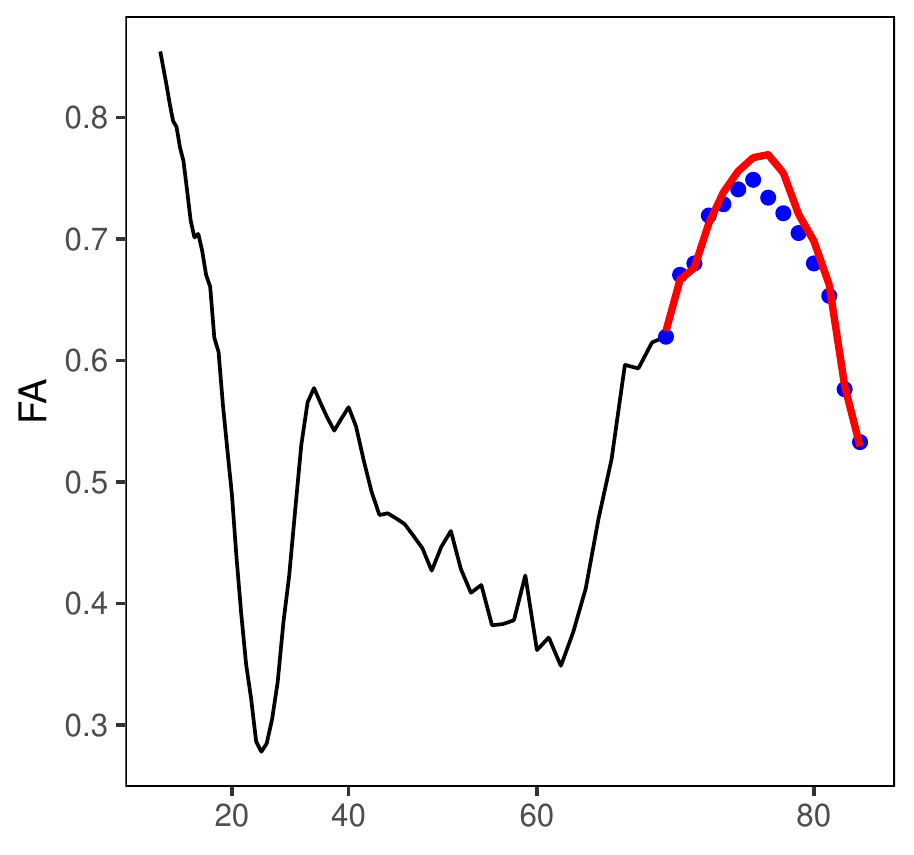}& \hspace*{0.7cm}
	\includegraphics[height=4.5cm,width=5.5cm]{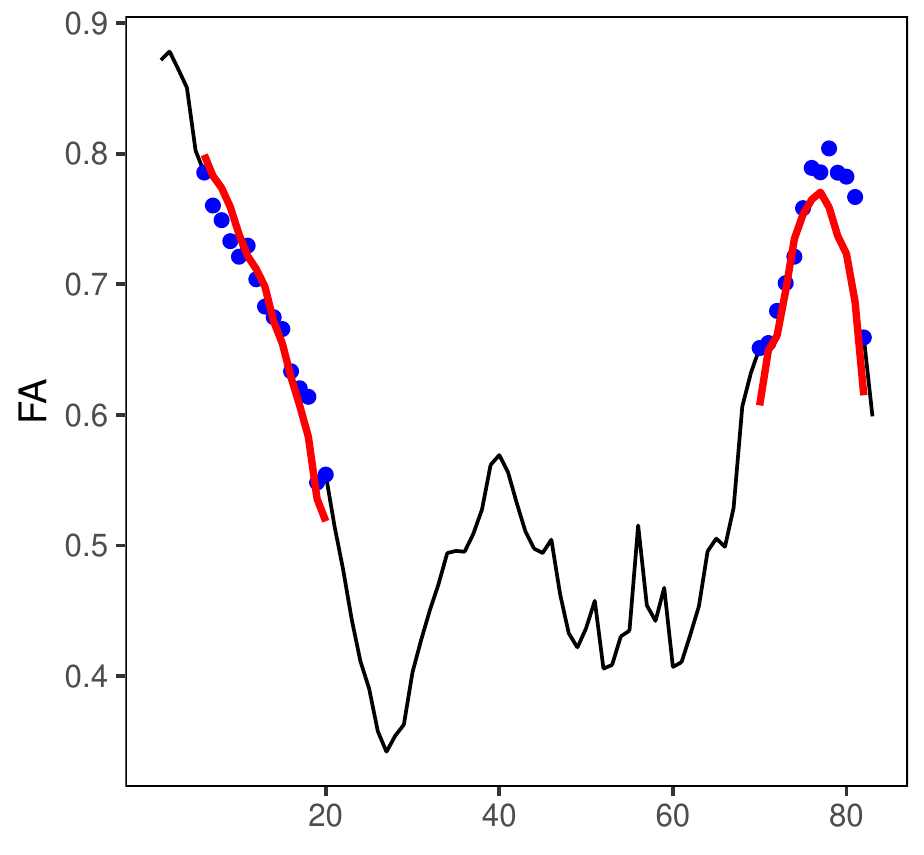}
	\end{array}$
	\caption{\textmd{Predicted profiles for four randomly chosen FA curves with different missing parts with $(a_1,a_2)=(1.5,0.2)$. Missing parts of trajectories from left to right and top to down: missing in left side, middle side, right side, both left and right side. Blue point: real data point; Red line: predicted profile}}
	\label{comfa} 
\end{figure*}

Next, we focus on the problem of recovering the missing parts $X_{iM_i}$ of $X_i$. Assume that the infinite-dimensional process $X_i$ is well approximated by the projection onto the function space $L^2(\mathcal{T})$ via the first $m$ eigenfunctions (\cite{yao2005afunctional}). In practice, the prediction for the trajectory $X_i(t)$ of the $i$th subject using the first $m$ eigenfunctions given in Section \ref{wme} can be approached by 
\begin{align*}
\hat{X}_i(t) = \hat{\mu}^{\text{NME}}(t) + \sum_{k=1}^{m}\hat{U}^{(\rho)}_{ij}\hat{\phi}^{\text{NME}}_j(t). 
\end{align*}
We randomly select four FA curves with different missing parts. The predicted profiles for these four curves are presented in Figure \ref{comfa}, showing that the predicted profiles are close to the real part. This demonstrate the ``NME" method by recovering the missing parts of incomplete trajectories encourages a better estimator comparing with the ``SUB" method with deleting them directly.

\section{Discussion}

In this paper, we address the problem for getting estimators of $\gamma$ in \eqref{mainmodel} with partially observed trajectories without and with measurement error. Basic elements of our approach are estimators of FPC scores for each partially observed trajectory. Specially, in the scenario that incomplete functional predictors observed without measurement error, we achieve it by modeling FPC scores of the missing part as linear functionals of the observed part of that trajectory. In the scenario where incomplete functional data is observed with measurement error, we obtain estimators of FPC scores via conditional expectation. Rates of convergence of the proposed estimators $\hat{\gamma}^{\text{NME}}$, $\hat{\gamma}^{\text{WME}}$ under different scenarios are established. We also compare the proposed methods with the ``SUB" or ``IN" method. We conclude from simulation studies that both the ``NME" and ``WME" methods borrowing strength from entire samples to get estimators of $\gamma$ in model \eqref{mainmodel} perform well in practice.

The methods proposed here can be extended to other models in terms of functional regression with partially observed trajectories, such as partial functional linear regression (see \cite{shin2009partial}). The framework established in this paper is based on the assumption that missing parts of trajectories are missing completely at random. In a number of applications, it is common to encounter that the underlying missing mechanism for dataset depends on systematic strategies (\cite{liebl2019csda}) that clearly violate MCAR assumption. Extension to this scenario is also of interest and significance in practice.

\section*{Appendix}
\begin{Lem}(\cite{kraus2015components}, Proposition 1.)\label{lem1}
	
	a) Let $\mathrm{E}\|X\|^2<\infty$ and assumption (A1) be satisfied. Then $\mathrm E(||\hat \mu^{\mathrm{NME}}-\mu||^2)=O(n^{-1})$ for $n\rightarrow\infty$.\par
	b) Let $\mathrm{E}\|X\|^4<\infty$ and observation pattern (A2) holds. Then $\mathrm E(||\hat C^{\mathrm{NME}}_X-C_X||_S^2)=O(n^{-1})$ for $n\rightarrow\infty$
	(here $||\cdot||_S$ denotes the Hilbert-Schmidt norm).	
\end{Lem}

\begin{Lem}(\cite{kneip2020optimal}, Theorem 4.1.) \label{lem2}
	
	Under the assumptions (B1)-(B5), we have that\\
	(a) $\mathrm{sup}_{t\in\mathcal{T}}|\hat{\mu}^{\mathrm{WME}}(t) - \mu(t)| = O_p(r_{\mu})$ with $r_{\mu}=h^2_{\mu} + 1/\sqrt{nNh_{\mu}} + 1/\sqrt{n}$.\\
	(b) $\mathrm{sup}_{(s,t)\in\mathcal{T}^2}|\hat{c}^{\mathrm{WME}}(s,t)-c_X(s,t)| = O_p(r_{\mu}+r_{c})$ with $r_{c} = h^2_{c} + 1/\sqrt{nMh^2_{c}} + 1/\sqrt{n}$.
\end{Lem}

\textbf{Proof of Theorem 3.1.}
The following results can be derived from the theory developed by \cite{bhatia1983perturbation}: 
\begin{align*}
\text{sup}_{j\geq 1} |\hat\lambda_j^{\mathrm{NME}} - \lambda_j| \leq \|\hat{C}^{\mathrm{NME}}_X - C_X\|, \,\,\,\,\,\,\,\, 
\text{sup}_{j\geq 1} \delta_j\|\hat{\phi_j}^{\mathrm{NME}} - \phi_j\| \leq 8^{1/2} \|\hat{C}^{\mathrm{NME}}_X - C_X\|.  \tag{13} \label{eigenre}
\end{align*}
Therefore, we obtain from Lemma \ref{lem1},
\begin{align*}
&\text{sup}_{j\geq 1} |\hat\lambda^{\mathrm{NME}}_j - \lambda_j| = O_p(n^{-1/2}),\\
&\text{sup}_{j\geq 1}\delta_j\|\hat{\phi}^{\mathrm{NME}}_j - \phi_j\| = O_p(n^{-1/2}).
\tag{14} \label{eigenvalue}
\end{align*}

Note that, 
\begin{align*}
\int_{\mathcal{T}} &(\hat{\gamma}^{\mathrm{NME}}(s)-\gamma(s))^2\text{ds}\\
= &\int_{\mathcal{T}} \left\{\sum_{j=1}^{m-1}\left[\frac{n^{-1}\sum_{i=1}^{n}[Y_i\hat{U}^{\mathrm{NME}}_{ij}]}{\hat{\lambda}^{\mathrm{NME}}_j}\hat{\phi}^{\mathrm{NME}}_j(s)-\frac{\text{E}[YU_j]}{\lambda_j}\phi_j(s) \right] \right\}^2\text{ds}\\
& + \int_{\mathcal{T}} \left\{\sum_{j=m}^{\infty}\frac{\text{E}[YU_j]}{\lambda_j}\phi_j(s)  \right\}^2\text{ds}\\
&+2\int_{\mathcal{T}}\left\{\sum_{j=1}^{m-1}\left[\frac{n^{-1}\sum_{i=1}^{n}[Y_i\hat{U}^{\mathrm{NME}}_{ij}]}{\hat{\lambda}^{\mathrm{NME}}_j}\hat{\phi}^{\mathrm{NME}}_j(s)-\frac{\text{E}[YU_j]}{\lambda_j}\phi_j(s) \right] \right\}
\left\{\sum_{j=m}^{\infty}\frac{\text{E}[YU_j]}{\lambda_j}\phi_j(s)\right\} \text{ds}\\
:=& A_1(n) + A_2(n) +A_3(n). \tag{15} \label{app1}
\end{align*}

For simplicity, we suppress the notation on ``NME". Assumption (A6) implies that $A_2(n)\rightarrow 0$ as $m\rightarrow\infty$. For $A_3(n)$, Cauchy-Schwarz inequality implies that $A^2_3(n)\leq A^2_1(n) \times A^2_2(n)\overset{p}{\rightarrow}0$. Combing the result \eqref{eigenvalue}, and the formula \eqref{app1}, we see that the result of the theorem follows if we can get the convergence rate of $\hat{U}_{ij}$ of the trajectories per subject with $\hat{U}_{ij} = \hat{U}_{ijO_i} + \hat U^{(\alpha)}_{ijM_i}$.

Denote the estimates of $U_{ijM_i}$, $C_{O_iO_i}$, $C_{O_iM_i}$, $\phi_{jM_i}$ as $\hat U_{ijM_i(-i)}$, $\hat C_{O_iO_i(-i)}$, $\hat C_{O_iM_i(-i)}$, $\hat \phi_{jM_i(-i)}$ with deleting the $i$th curves $X_i(t)$. Let $\tilde{\xi}^{(\rho)}_{ijM_i} = (C^{(\rho)}_{O_iO_i})^{-1}C_{O_iM_i}\phi_{jM_i}$ with $C^{(\rho)}_{O_iO_i} = C_{O_iO_i}+\rho \mathcal{F}_{O_i}$, $\tilde{U}_{ijM_i}^{(\rho)} = \langle \tilde{\xi}^{(\rho)}_{ijM_i}, X_{iO_i} \rangle$, and the notation $\tilde{\xi}_{ijM_i}$ $\tilde{U}_{ijM_i}$ are corresponded to the symbols $\tilde{\xi}^{(\rho)}_{ijM_i}$,$\tilde{U}_{ijM_i}^{(\rho)}$ with $\rho=0$. Since
\begin{align*}
\text E\|\hat U_{ijM_i}^{(\rho)}-\tilde{U}_{ijM_i}\|^2
=&~~ \text E\|\hat U_{ijM_i}^{(\rho)}-\tilde{U}_{ijM_i}^{(\rho)}+\tilde{U}_{ijM_i}^{(\rho)}-\tilde{U}_{ijM_i}\|^2\\
=&~~ 2\text E\|\hat U_{ijM_i}^{(\rho)}-\tilde{U}_{ijM_i}^{(\rho)}\|^2+ 2\|\tilde{U}_{ijM_i}^{(\rho)}-\tilde{U}_{ijM_i}\|^2\\
\leq &~~ 4\text E\|\hat U_{ijM_i}^{(\rho)}-\hat{U}_{ijM_i(-i)}^{(\rho)}\|^2 + 4\text E\|\hat U_{ijM_i(-i)}^{(\rho)}-\tilde{U}_{ijM_i}^{(\rho)}\|^2 \\
& + 2\|\tilde{U}_{ijM_i}^{(\rho)}-\tilde{U}_{ijM_i}\|^2,
\tag{16} \label{scoresep}
\end{align*}
we then analyze the terms $\text E\|\hat U_{ijM_i}^{(\rho)}-\hat{U}_{ijM_i(-i)}^{(\rho)}\|^2$,
$\text E\|\hat U_{ijM_i(-i)}^{(\rho)}-\tilde{U}_{ijM_i}^{(\rho)}\|^2$,
$\|\tilde{U}_{ijM_i}^{(\rho)}-\tilde{U}_{ijM_i}\|^2$ in turn.
Let $\hat\xi_{ijM_i(-i)}^{(\rho)} = (\hat C^{(\rho)}_{O_iO_i(-i)})^{-1}\hat C_{O_iM_i(-i)}\hat\phi_{jM_i(-i)}$. Then
\begin{align*}
\text E\|\hat U_{ijM_i(-i)}^{(\rho)}-\tilde{U}_{ijM_i}^{(\rho)}\|^2
=&~~ \text E\langle\hat\xi_{ijM_i(-i)}^{(\rho)}-\tilde{\xi}_{ijM_i}^{(\rho)},X_{iO_i}\rangle^2\\
=&~~ \text E\{\text E[\langle\hat\xi_{ijM_i(-i)}^{(\rho)}-\tilde{\xi}_{ijM_i}^{(\rho)},X_{iO_i}\rangle^2| \{X_{kO_i}, k\neq i\}]\}\\
=&~~ \text E||C_{O_iO_i}^{1/2}((\hat C^{(\rho)}_{O_iO_i(-i)})^{-1}\hat C_{O_iM_i(-i)}\hat\phi_{jM_i(-i)}-( C^{(\rho)}_{O_iO_i})^{-1}C_{O_iM_i}\phi_{jM_i})||^2\\
\leq &~~ 4\left\{\text E ||C_{O_iO_i}^{1/2}(\hat C^{(\rho)}_{O_iO_i(-i)})^{-1} (\hat C_{O_iM_i(-i)}-C_{O_iM_i})(\hat\phi_{jM_i(-i)} -\phi_{jM_i})||^2 \right.\\
& + \text E ||C_{O_iO_i}^{1/2}(\hat C^{(\rho)}_{O_iO_i(-i)})^{-1} C_{O_iM_i}(\hat\phi_{jM_i(-i)}-\phi_{jM_i})||^2\\
& + \text E ||C_{O_iO_i}^{1/2}(\hat C^{(\rho)}_{O_iO_i(-i)})^{-1}(\hat C_{O_iM_i(-i)}-C_{O_iM_i})\phi_{jM_i}||^2\\
& \left.+ \text E ||C_{O_iO_i}^{1/2}((\hat C^{(\rho)}_{O_iO_i(-i)})^{-1} - (C^{(\rho)}_{O_iO_i})^{-1}) C_{O_iM_i}\phi_{jM_i}||^2 \right\}\\
:=& B_1+B_2+B_3+B_4.
\tag{17} \label{scoreridge}
\end{align*}

Let $\mathcal{F}_m=\{\frac{\lambda_m}{2}<\hat\lambda_m<\frac{3}{2}\lambda_m\}$.  Suppose the event $\mathcal{F}_m$
holds. Otherwise, we have 
$\text{P}(|\hat\lambda_m-\lambda_m|\geq\frac{\lambda_m}{2})\leq \text{P}(\|\hat{C}^{\mathrm{NME}}_X - C_X\|\geq \frac{\lambda_m}{2})\rightarrow 0$ from assumption (A4).
We have the following results for terms $B_1$ to $B_4$ with the equality
\begin{align*}\begin{split}
\left(\hat C^{(\rho)}_{O_iO_i(-i)}\right)^{-1}-\left(C^{(\rho)}_{O_iO_i}\right)^{-1}
=(\hat C_{O_iO_i(-i)}-C_{O_iO_i})\left( C^{(\rho)}_{O_iO_i}\right)^{-1}\left(\hat C^{(\rho)}_{O_iO_i(-i)}\right)^{-1}.
\end{split}\end{align*}
For the term $B_1$, 
\begin{align*}\begin{split}
B_1
&\leq \text E [\|C_{O_iO_i}^{1/2}\|_2^2 \cdot \|(\hat C^{(\rho)}_{O_iO_i(-i)})^{-1}\|_\infty^2 \cdot \|\hat C_{O_iM_i(-i)}-C_{O_iM_i}\|_2^2 \cdot \|\hat\phi_{jM_i(-i)}-\phi_{jM_i}\|^2]\\
&=O(n^{-2}\delta^{-2}_j)\cdot O(\rho^{-2}).
\end{split}\end{align*}
Denote $\|\cdot\|_\infty$ as the operator norm. For the term $B_2$, under the assumption (A7), $\text{E} \|C_{O_iO_i}^{1/2}\|_\infty^2<\infty$ and the result \eqref{eigenvalue}, it is clear that
\begin{align*}\begin{split}
B_2
&\leq \text E [\|C_{O_iO_i}^{1/2}\|_\infty^2 \cdot\|(\hat C^{(\rho)}_{O_iO_i(-i)})^{-1}C_{O_iM_i}\|_2^2 \cdot \|\hat\phi_{jM_i(-i)}-\phi_{jM_i}\|^2]\\
&\leq \sum_j\sum_k \frac{r_{M_iO_ijk}^2 }{(\lambda_{O_i O_i k}+\rho)^2}\cdot O(n^{-1}\delta^{-2}_j) = O(n^{-1}\delta^{-2}_j).
\end{split}\end{align*}
For the term $B_3$,
\begin{align*}\begin{split}
B_3
&\leq \text E [\|C_{O_iO_i}^{1/2}\|_2^2 \cdot\|(\hat C^{(\rho)}_{O_iO_i(-i)})^{-1}\|_\infty^2 \cdot \|\hat C_{O_iM_i(-i)}-C_{O_iM_i}\|_2^2 \cdot \|\phi_{jM_i}\|^2]\\
&= O(n^{-1}\rho^{-2}).
\end{split}\end{align*}
Note that $\frac{\rho\lambda_{O_iO_ik}}{(\lambda_{O_iO_ik}+\rho)^2}<1$. Under the assumption (A7), we have that
\begin{align*}\begin{split}
B_4
&\leq \text E \left[\|C_{O_iO_i}^{1/2} \cdot (C^{(\rho)}_{O_iO_i})^{-1}\cdot (\hat C^{(\rho)}_{O_iO_i(-i)})^{-1} \cdot C_{O_iM_i}\|_2^2 \cdot \|\hat C_{O_iO_i(-i)}-C_{O_iO_i}\|_2^2
\cdot \|\phi_{jM_i}\|^2\right]\\
&\leq \left\{\sum_j\sum_k\frac{\rho\lambda_{O_iO_ik}}{(\lambda_{O_iO_ik}+\rho)^2}\cdot \frac{{r_{O_iM_ijk}}^2}{(\lambda_{O_iO_ik}+\rho)^2}\cdot\rho^{-1}\right\}\cdot O(n^{-1})\\
&= O(n^{-1})\cdot O(\rho^{-1}).
\end{split}\end{align*}
These results combined with \eqref{scoreridge} indicate
\begin{equation*}
\text E\|\hat U_{ijM_i(-i)}^{(\rho)}-\tilde{U}_{ijM_i}^{(\rho)}\|^2 = O(n^{-1}\rho^{-2}+n^{-1}\delta^{-2}_j). \tag{18} \label{scoreapp}
\end{equation*}
We then analyze $\text E\|\hat U_{ijM_i}^{(\rho)}-\hat{U}_{ijM_i(-i)}^{(\rho)}\|^2$,
\begin{align*}
\text E\|\hat U_{ijM_i}^{(\rho)}-\hat{U}_{ijM_i(-i)}^{(\rho)}\|
=&~~ \text E\langle\hat\xi_{ijM_i}^{(\rho)}-\hat{\xi}_{ijM_i(-i)}^{(\rho)},X_{iO_i}\rangle\\
\leq &~~ \{\text E\|\hat\xi_{ijM_i}^{(\rho)}-\hat{\xi}_{ijM_i(-i)}^{(\rho)}\|^2\}^{1/2} \{\text E\| X_{iO_i} \|^2\}^{1/2}\\
\leq &~~L \{\text E\|\hat\xi_{ijM_i}^{(\rho)}-\hat{\xi}_{ijM_i(-i)}^{(\rho)}\|^2\}^{1/2}, \tag{19} \label{ualpha}
\end{align*}
where the last inequality holds from the finite second moment of $X$ that is bounded by constant $L$.
We also have, 
\begin{align*}
\text E \|\hat\xi_{ijM_i}^{(\rho)}-\hat{\xi}_{ijM_i(-i)}^{(\rho)}\|^2
= &~~ \text E \| \left((\hat C^{(\rho)}_{O_iO_i})^{-1}\hat C_{O_iM_i} - (\hat C^{(\rho)}_{O_iO_i(-i)})^{-1}\hat C_{O_iM_i(-i)}\right) \hat\phi_{jM_i(-i)} \|^2\\
=&~~ \text E\| \left[\left((\hat C^{(\rho)}_{O_iO_i})^{-1}-(\hat C^{(\rho)}_{O_iO_i(-i)})^{-1}\right)\hat C_{O_iM_i}\right. \\
& +\left. (\hat C^{(\rho)}_{O_iO_i(-i)})^{-1}(\hat C_{O_iM_i}-\hat C_{O_iM_i(-i)})\right]\hat\phi_{jM_i(-i)} \|^2\\
\leq & ~~2 \left\{\text E\| \left((\hat C^{(\rho)}_{O_iO_i})^{-1}-(\hat C^{(\rho)}_{O_iO_i(-i)})^{-1}\right)\hat C_{O_iM_i}\|^2 \right. \\
&+  \left.\text E \|(\hat C^{(\rho)}_{O_iO_i(-i)})^{-1}(\hat C_{O_iM_i}-\hat C_{O_iM_i(-i)})\|^2 \right.\}.
\tag{20} \label{scoredelet}
\end{align*}
Note that
$$
\text E \|\hat C_{O_iM_i}-\hat C_{O_iM_i(-i)}\|^2 = O(n^{-2}),
$$
$$
\text E\|\left((\hat C^{(\rho)}_{O_iO_i})^{-1}-(\hat C^{(\rho)}_{O_iO_i(-i)})^{-1}\right)\hat C_{O_iM_i}\|^2 = O(n^{-2}),
$$
$$
\text E \|(\hat C^{(\rho)}_{O_iO_i(-i)})^{-1}(\hat C_{O_iM_i}-\hat C_{O_iM_i(-i)})\|^2 = O(n^{-2}\rho^{-2}).
$$
Combining formulas \eqref{ualpha} and \eqref{scoredelet}, we deduce that
\begin{equation*}
\text E\|\hat U_{ijM_i}^{(\rho)}-\hat{U}_{ijM_i(-i)}^{(\rho)}\|^2 = O(n^{-2}\rho^{-2}). \tag{21} \label{xipro}
\end{equation*}
On the other hand, 
\begin{equation*}
\text E \parallel \tilde{U}_{ijM_i}^{(\rho)}-\tilde{U}_{ijM_i}\parallel^2=O(\rho), \tag{22} \label{xscoreapprorate}
\end{equation*}
\begin{align*}
\text{var}(\tilde{U}_{ijM_i}-U_{ijM_i}) &= \langle\phi_{jM_i},C_{M_iM_i}\phi_{jM_i}\rangle - \langle\phi_{jM_i}, C_{M_iO_i}C^{-1}_{O_iO_i}C_{O_iM_i}\phi_{jM_i}\rangle \\
&:=V_{ij}. \tag{23}\label{uvar}
\end{align*}
Therefore, with $n\rho^3\rightarrow 0$ and the formulas \eqref{scoresep}, \eqref{scoreapp}, \eqref{xipro}-\eqref{uvar}, we have that
\begin{equation*}
\text E \|\hat U_{ijM_i}^{(\rho)}-U_{ijM_i}\|^2=O(n^{-1}\rho^{-2} + n^{-1}\delta^{-2}_j + V_{ij}).
\end{equation*}
Then the result is proved with $n\rho^3\rightarrow 0$.

\textbf{Proof of Theorem 3.2.} Let $\bm{\tilde{U}}_{i} = (\tilde{U}_{i1},\cdots, \tilde{U}_{im})^T$, $\bm{{U}}_{i} = ({U}_{i1},\cdots, {U}_{im})^T$. The covariance matrix of $\bm{\tilde{U}}_{i}$ is $\text{var}(\bm{{U}}_{i})=\Xi\mathbf{\Sigma}^{-1}_{{\mathbf{Z}}_i}\Xi^T$ with $\Xi = \text{cov}(\bm{\tilde{U}}_{i},{\mathbf{Z}_i})=(\lambda_1\mathbf{{\bm\phi}}_{i1},\cdots,\lambda_m\mathbf{{\bm\phi}}_{im})^T$. Moreover, $\text{var}(\bm{\tilde{U}}_{i}-\bm{{U}}_{i}) = \mathbf{\Lambda}-\Xi\mathbf{\Sigma_{{\mathbf{Z}}_i}}\Xi^T$. Combining these results with formulas \eqref{eigenvalue}, \eqref{pace} and the results of Lemma \ref{lem2}, the result of Theorem 3 is obtained by replacing $\hat{U}^{\text{NME}}_{ij}$ with $\hat{U}^{\text{WME}}_{ij}$ in \eqref{app1} with assumptions (B1)-(B6).

\bibliographystyle{plain}

\bibliography{FLM_missing.bib}

\begin{thebibliography}{10}

\bibitem{besse1986principal}
Philippe Besse and James~O Ramsay.
\newblock Principal components analysis of sampled functions.
\newblock {\em Psychometrika}, 51(2):285--311, 1986.

\bibitem{bhatia1983perturbation}
Rajendra Bhatia, Chandler Davis, and Alan McIntosh.
\newblock Perturbation of spectral subspaces and solution of linear operator
  equations.
\newblock {\em Linear Algebra and its Applications}, 52:45--67, 1983.

\bibitem{cardot1999functional}
Herv{\'e} Cardot, Fr{\'e}d{\'e}ric Ferraty, and Pascal Sarda.
\newblock Functional linear model.
\newblock {\em Statistics \& Probability Letters}, 45(1):11--22, 1999.

\bibitem{che2017trajectory}
Menglu Che, Linglong Kong, Rhonda~C Bell, and Yan Yuan.
\newblock Trajectory modeling of gestational weight: A functional principal
  component analysis approach.
\newblock {\em PloS one}, 12(10):e0186761, 2017.

\bibitem{crambes2009smoothing}
Christophe Crambes, Alois Kneip, and Pascal Sarda.
\newblock Smoothing splines estimators for functional linear regression.
\newblock {\em The Annals of Statistics}, 37(1):35--72, 2009.

\bibitem{delaigle2016approximating}
A~Delaigle and P~Hall.
\newblock Approximating fragmented functional data by segments of markov
  chains.
\newblock {\em Biometrika}, 103(4):779--799, 2016.

\bibitem{goldberg2014predicting}
Yair Goldberg, Ya’acov Ritov, and Avishai Mandelbaum.
\newblock Predicting the continuation of a function with applications to call
  center data.
\newblock {\em Journal of Statistical Planning and Inference}, 147:53--65,
  2014.

\bibitem{hall2007methodology}
Peter Hall and Joel~L Horowitz.
\newblock Methodology and convergence rates for functional linear regression.
\newblock {\em The Annals of Statistics}, 35(1):70--91, 2007.

\bibitem{hall2006properties}
Peter Hall, Hans-Georg M{\"u}ller, and Jane-Ling Wang.
\newblock Properties of principal component methods for functional and
  longitudinal data analysis.
\newblock {\em The Annals of Statistics}, pages 1493--1517, 2006.

\bibitem{hansen1990discrete}
Per~Christian Hansen.
\newblock The discrete picard condition for discrete ill-posed problems.
\newblock {\em BIT Numerical Mathematics}, 30(4):658--672, 1990.

\bibitem{horvath2012inference}
Lajos Horv{\'a}th and Piotr Kokoszka.
\newblock {\em Inference for functional data with applications}, volume 200.
\newblock Springer Science \& Business Media, 2012.

\bibitem{james2000principal}
Gareth~M James, Trevor~J Hastie, and Catherine~A Sugar.
\newblock Principal component models for sparse functional data.
\newblock {\em Biometrika}, 87(3):587--602, 2000.

\bibitem{kneip2020optimal}
Alois Kneip and Dominik Liebl.
\newblock On the optimal reconstruction of partially observed functional data.
\newblock {\em Annals of Statistics}, 48(3):1692--1717, 2020.

\bibitem{kraus2015components}
David Kraus.
\newblock Components and completion of partially observed functional data.
\newblock {\em Journal of the Royal Statistical Society: Series B (Statistical
  Methodology)}, 77(4):777--801, 2015.

\bibitem{li2010uniform}
Yehua Li and Tailen Hsing.
\newblock Uniform convergence rates for nonparametric regression and principal
  component analysis in functional/longitudinal data.
\newblock {\em The Annals of Statistics}, 38(6):3321--3351, 2010.

\bibitem{liebl2013modeling}
Dominik Liebl.
\newblock Modeling and forecasting electricity spot prices: A functional data
  perspective.
\newblock {\em The Annals of Applied Statistics}, 7(3):1562--1592, 2013.

\bibitem{liebl2019csda}
Dominik Liebl and Stefan Rameseder.
\newblock Partially observed functional data: The case of systematically
  missing parts.
\newblock {\em Computational Statistics and Data Analysis}, 131:104--115, 2019.

\bibitem{marx1999generalized}
Brian~D Marx and Paul~HC Eilers.
\newblock Generalized linear regression on sampled signals and curves: a
  p-spline approach.
\newblock {\em Technometrics}, 41(1):1--13, 1999.

\bibitem{morris2015functional}
Jeffrey~S Morris.
\newblock Functional regression.
\newblock {\em Annual Review of Statistics and Its Application}, 2:321--359,
  2015.

\bibitem{ramsay2005functional}
James Ramsay.
\newblock Functional data analysis.
\newblock {\em Encyclopedia of Statistics in Behavioral Science}, 2005.

\bibitem{ramsay1991some}
James~O Ramsay and CJ~Dalzell.
\newblock Some tools for functional data analysis.
\newblock {\em Journal of the Royal Statistical Society: Series B
  (Methodological)}, 53(3):539--561, 1991.

\bibitem{reiss2017methods}
Philip~T Reiss, Jeff Goldsmith, Han~Lin Shang, and R~Todd Ogden.
\newblock Methods for scalar-on-function regression.
\newblock {\em International Statistical Review}, 85(2):228--249, 2017.

\bibitem{rice1991estimating}
John~A Rice and Bernard~W Silverman.
\newblock Estimating the mean and covariance structure nonparametrically when
  the data are curves.
\newblock {\em Journal of the Royal Statistical Society: Series B
  (Methodological)}, 53(1):233--243, 1991.

\bibitem{riesz1955b}
F~Riesz and Sz~Nagy.
\newblock B.(1990). functional analysis.
\newblock {\em Dover Publications, Inc., New York. First published in},
  3(6):35, 1955.

\bibitem{shang2014survey}
Han~Lin Shang.
\newblock A survey of functional principal component analysis.
\newblock {\em AStA Advances in Statistical Analysis}, 98(2):121--142, 2014.

\bibitem{shin2009partial}
Hyejin Shin.
\newblock Partial functional linear regression.
\newblock {\em Journal of Statistical Planning and Inference},
  139(10):3405--3418, 2009.

\bibitem{staniswalis1998nonparametric}
Joan~G Staniswalis and J~Jack Lee.
\newblock Nonparametric regression analysis of longitudinal data.
\newblock {\em Journal of the American Statistical Association},
  93(444):1403--1418, 1998.

\bibitem{wang2019wavelet}
Yafei Wang, Linglong Kong, Bei Jiang, Xingcai Zhou, Shimei Yu, Li~Zhang, and
  Giseon Heo.
\newblock Wavelet-based lasso in functional linear quantile regression.
\newblock {\em Journal of Statistical Computation and Simulation},
  89(6):1111--1130, 2019.

\bibitem{yao2005afunctional}
Fang Yao, Hans-Georg M{\"u}ller, and Jane-Ling Wang.
\newblock Functional data analysis for sparse longitudinal data.
\newblock {\em Journal of the American Statistical Association},
  100(470):577--590, 2005.

\bibitem{yao2005bfunctional}
Fang Yao, Hans-Georg M{\"u}ller, and Jane-Ling Wang.
\newblock Functional linear regression analysis for longitudinal data.
\newblock {\em The Annals of Statistics}, pages 2873--2903, 2005.

\bibitem{yu2016partial}
Dengdeng Yu, Linglong Kong, and Ivan Mizera.
\newblock Partial functional linear quantile regression for neuroimaging data
  analysis.
\newblock {\em Neurocomputing}, 195:74--87, 2016.

\bibitem{zhao2012wavelet}
Yihong Zhao, R~Todd Ogden, and Philip~T Reiss.
\newblock Wavelet-based lasso in functional linear regression.
\newblock {\em Journal of Computational and Graphical Statistics},
  21(3):600--617, 2012.

\bibitem{zhu2012multivariate}
Hongtu Zhu, Runze Li, and Linglong Kong.
\newblock Multivariate varying coefficient model for functional responses.
\newblock {\em Annals of Statistics}, 40(5):2634, 2012.

\end{thebibliography}

\end{document}